\documentclass{iaus}
\usepackage{graphicx}

\title[Magnetically dominated cooling towers] 
{Modelling magnetically dominated and radiatively cooling jets}

\author[Huarte-Espinosa, Frank \& Blackman]   
{Mart\'{\i}n Huarte-Espinosa$^{1,2}$, Adam Frank$^1$
 \and Eric Blackman$^1$}

\affiliation{$^1$Department of Physics and Astronomy, University of Rochester, \\
600 Wilson Boulevard, Rochester, NY, 14627-0171 \\ 
emails: {\tt martinhe; afrank; blackman @pas.rochester.edu}\\
$^2$Kavli Institute for Cosmology Cambridge, Madingley Road, \\ 
Cambridge CB3 0HA, UK}

\pubyear{2011}
\volume{275}  
\pagerange{?--?}
 \date{?? and in revised form ??}
\jname{Jets at all Scales}
\editors{A.C. Editor, B.D. Editor \& C.E. Editor, eds.}
\begin{document}

\maketitle

\begin{abstract}
Using 3D-MHD Eulerian-grid numerical simulations, we study the
formation and evolution of rising magnetic towers propagating into
an ambient medium.  The towers are generated from a localized
injection of pure magnetic energy.  No rotation is imposed on the
plasma.  We compare the evolution of a radiatively cooling tower
with an adiabatic one, and find that both bend due to pinch
instabilities.  Collimation is stronger in the radiative cooling
case; the adiabatic tower tends to expand radially. Structural
similarities are found between these towers and the millimeter scale
magnetic towers produced in laboratory experiments.
\end{abstract}

\firstsection 
\section{Introduction}

Non-relativistic jets are observed in the vicinities of several
Protostellar Objects, Young Stellar Objects (YSOs) and post-AGB
stars.  Plausible models suggest that jets are launched and collimated
by accretion, rotation and magnetic mechanisms in the ``central
engine'' (see \cite[Pudritz et al.  2007]{pudritz07}, for a review).
The relative extent over which magnetic energy dominates the outflow
kinetic energy in a propagating jet has traditionally divided them
into two classes: magnetocentrifugal (\cite[Blandford \& Payne
1982]{blandford82}; \cite[Ouyed \& Pudritz 1997]{ouyed97};~
\cite[Blackman et~al. 2001]{blackman01};~ \cite[Mohamed \& Podsiadlowski
2007]{mohamed07}), in which magnetic fields only dominate out to
the Alfv\'en radius, or Poynting flux dominated (PFD,~ \cite[Lynden-Bell
1996]{bell96}; \cite[Ustyugova et al.  2000]{ustyugova00};
\cite[Lovelace et al. 2002]{lovelace02}; \cite[Nakamura \& Meier
2004]{nakamura04}) in which magnetic fields dominate the jet
structure, acting as a magnetic piston over very large distances
from the engine.  PFD jets carry large electric currents along which
generate strong tightly wound helical magnetic fields around the
jet axis.  Simulations of such jets have found that these magnetic
fields play a role in  the formation of current-driven kink
instabilities and the stabilization of Kelvin-Helmholtz (KH) modes
in jets (e.g. see \cite[Nakamura \& Meier 2004]{nakamura04}).

The effects of plasma radiative-cooling have been followed only in
few simulations of magnetized, kinetic energy-dominated jets. These,
have been found to form both thin cocoons and nose cones (e.g.
\cite[Blondin et~al.  1990]{blondin90}; \cite[Frank et~al.
1998]{frank98}), and to be more susceptible to KH~instabilities
relative to adiabatic jets (\cite[Hardee \& Stone 1997]{hardee97},
and references therein).  Correlations between the structure of
kinetic energy-dominated jets and their power have been extensively
explored. There has been less work on PFD jets.

Magnetic fields with initially primarily poloidal (radial and
vertical) geometries anchored to accretion discs were shown to form
tall, highly wound and helical magnetic structures, or magnetic
towers, that expand vertically when laterally supported in pressure
equilibrium with the ambient gas (\cite[Lynden-Bell
1996,~2003]{bell96,bell03}).  The local injection of pure toroidal
magnetic energy, without imposing any rotation on the plasma, has
been shown to form magnetic towers in laboratory experiments
(\cite[Lebedev~et al.~2005]{lab1}. These experiments were modeled
with numerical simulations by \cite[Ciardi~et al.~2007] {lab2}) and
are analogous to 3D-MHD numerical simulations of AGN jets by \cite[Li
et~al. (2006,]{li06} and subsequent papers).  Here we use a modified
version of the implementation of \cite[Li et~al.]{li06}, in order
to study magnetically driven and radiatively-cooling jets, or
magnetic towers, motivated by the contexts of Protostellar Objects,
YSOs and Planetary Nebulae.

\section{Model}

We form magnetic towers using 3D-MHD Eulerian-grid numerical
simulations by locally injecting pure magnetic energy and compare
the evolution of these towers as they propagate into an ambient
``interstellar'' medium (ISM) for radiative cooling vs. adiabatic
cases.  No rotation is imposed on the plasma at the base.  The ISM
gas is modelled with an ideal gas equation of state, a ratio of
specific heats of $\gamma=\,$5$/$3, a uniform number density of
100\,cm$^{-3}$, a constant temperature of 10000\,K and null velocity.
We start the simulations with a helical magnetic field inside a
central cylinder with both radius and height of 50$\pi$\,AU. The
helical geometry of the initial magnetic field is described from
its vector potential (i.e.
${\bf B=\nabla \times  A}$) 
\begin{equation}
{\bf A}(r,z) = \left\{
   \begin{array}{c l}
          \frac{r}{4} (\cos{2\,r} + 1)( \cos{2\,z} + 1 ) \hat{\phi} + 
          \frac{\alpha}{8} (\cos{2\,r} + 1)( \cos{2\,z} + 1 ) \hat{k},
             & \mbox{for}~r,z < \pi/2; \\
          0, & \mbox{for}~r,z \ge \pi/2,
   \end{array} \right.
\label{apot} 
\end{equation}
\noindent in cylindrical coordinates. The parameter~$\alpha$  is
an integer with units of length and determines the ratio of toroidal
to poloidal magnetic fluxes, and electric currents.  For the initial
conditions $\alpha=\,3$, but $\alpha=\,15$ at all later times.

A cubic computational domain with 128$^3$ fixed cells is used.
Boundary conditions are set to periodic at both $x=\pm$800\,AU and
$y=\pm\,$800\,AU, to reflective at $z=\,$0, and to outflow at
$z=\,$1600\,AU. We use BlueHive\footnote{
https://www.rochester.edu/its/web/wiki/crc/index.php/BlueHive\_Cluster}, an
IBM parallel cluster of the Center for Research Computing of the
University of Rochester, to run simulations for about two~weeks,
using 32 processors.

Using the above initial and boundary conditions, we solve the
equations of radiative-magnetohydrodynamics in three-dimensions
with the AMR parallel code \textit{AstroBEAR}\footnote{
http://www.pas.rochester.edu/$\sim$bearclaw/} (\cite[Cunningham et
al.  2009]{astrobear}). A source term is implemented to the induction
equation to continuously inject magnetic fields via (\ref{apot}),
into the computational domain.  For numerical stabilization, static
gas is also injected into regions where ${\bf A}(r,z)$ is not null,
but its contribution to both mass and thermal energy  is negligible.
The ionization of both~H and~He, the chemistry of~H2 and optically
thin cooling are considered (in one of our simulations) using tables
of \cite[Dalgarno \& McCray (1972)]{dm}.  To focus on the minimalist
physics of magnetic tower expansion, no gravitational, viscous or
general-relativistic processes are considered. This implementation
extends the one of \cite[Li~et~al.  (2006)]{li06} to stellar scales
and to the radiative-MHD regime.

\section{Results and discussion}

Here we discuss some of our preliminary results.
Higher resolution simulations, further details and analyses
will be presented in Huarte~Espinosa, Frank \& Blackman~2011 (in prep.).

\begin{figure}
\begin{center}
\includegraphics[width=.24\columnwidth]{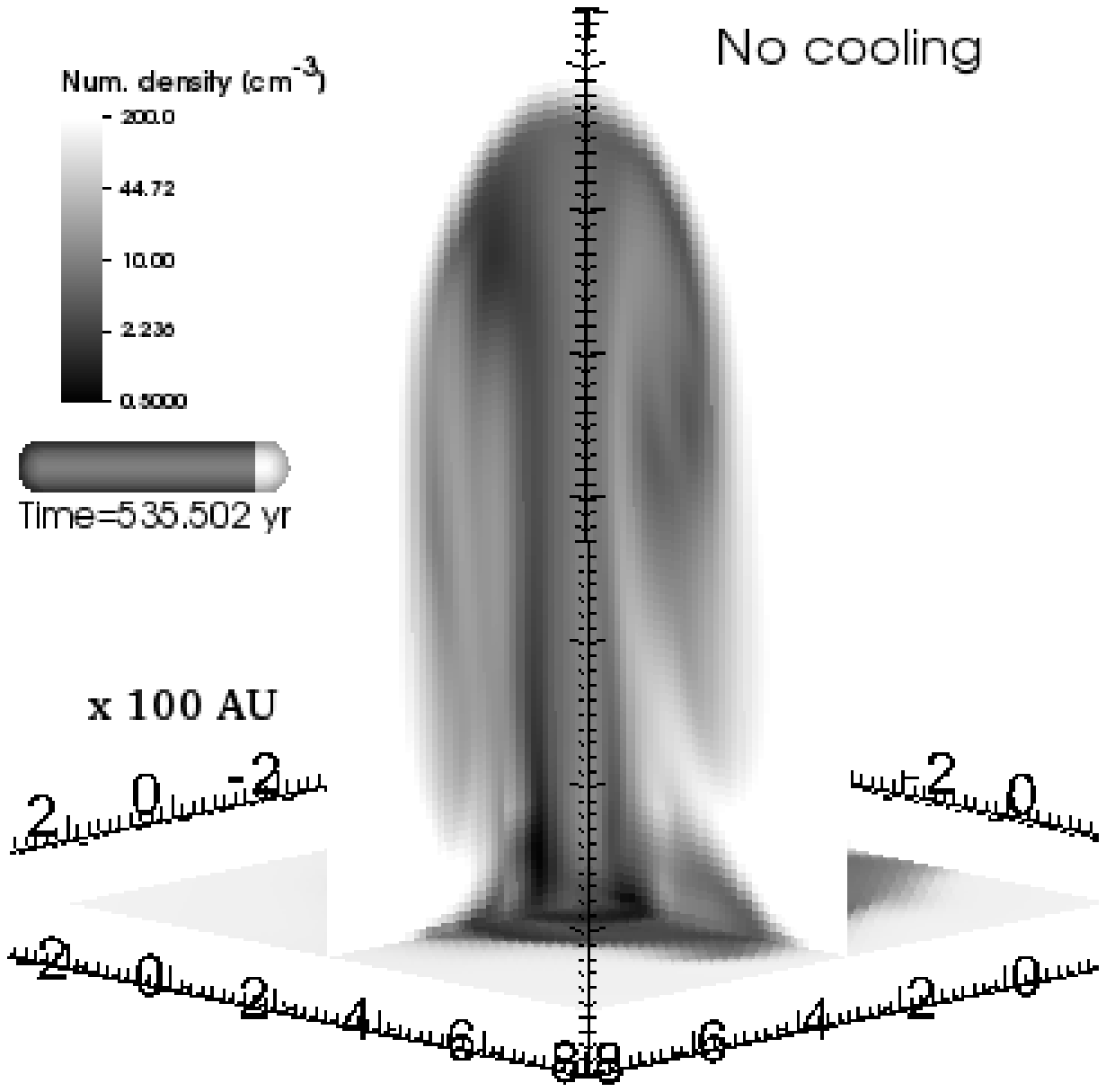} 
\includegraphics[width=.24\columnwidth]{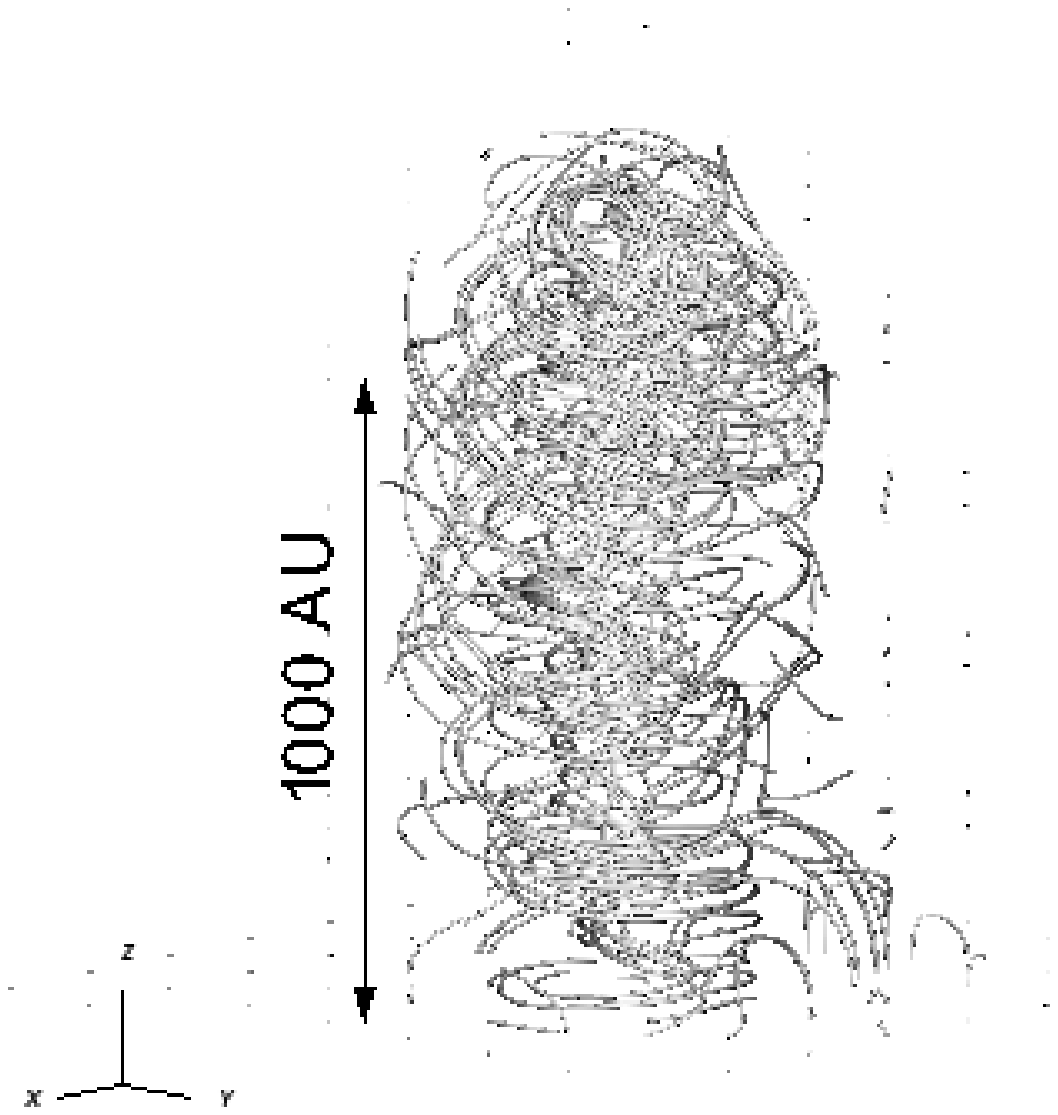} 
\includegraphics[width=.14\columnwidth]{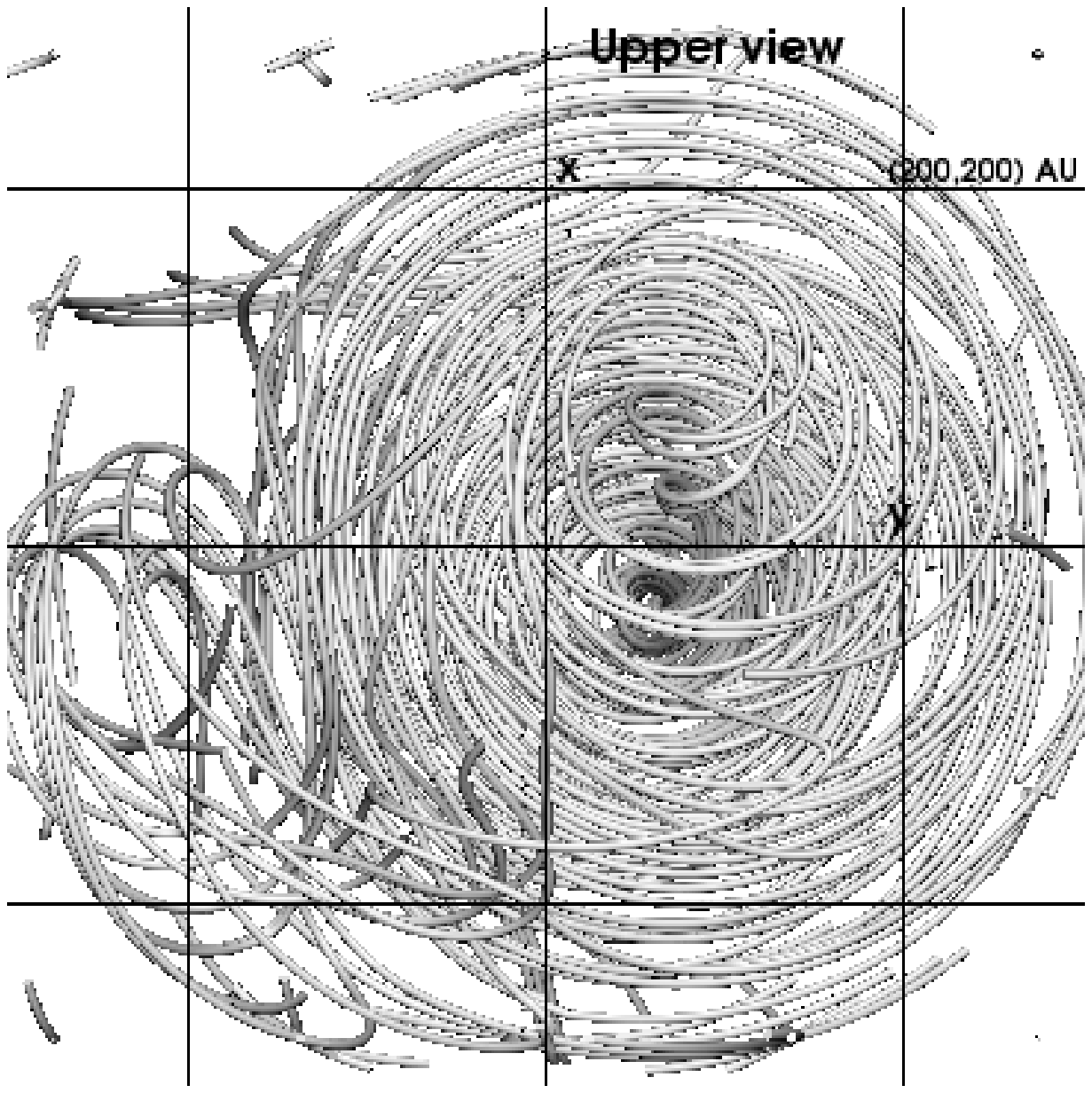} \\
\includegraphics[width=.24\columnwidth]{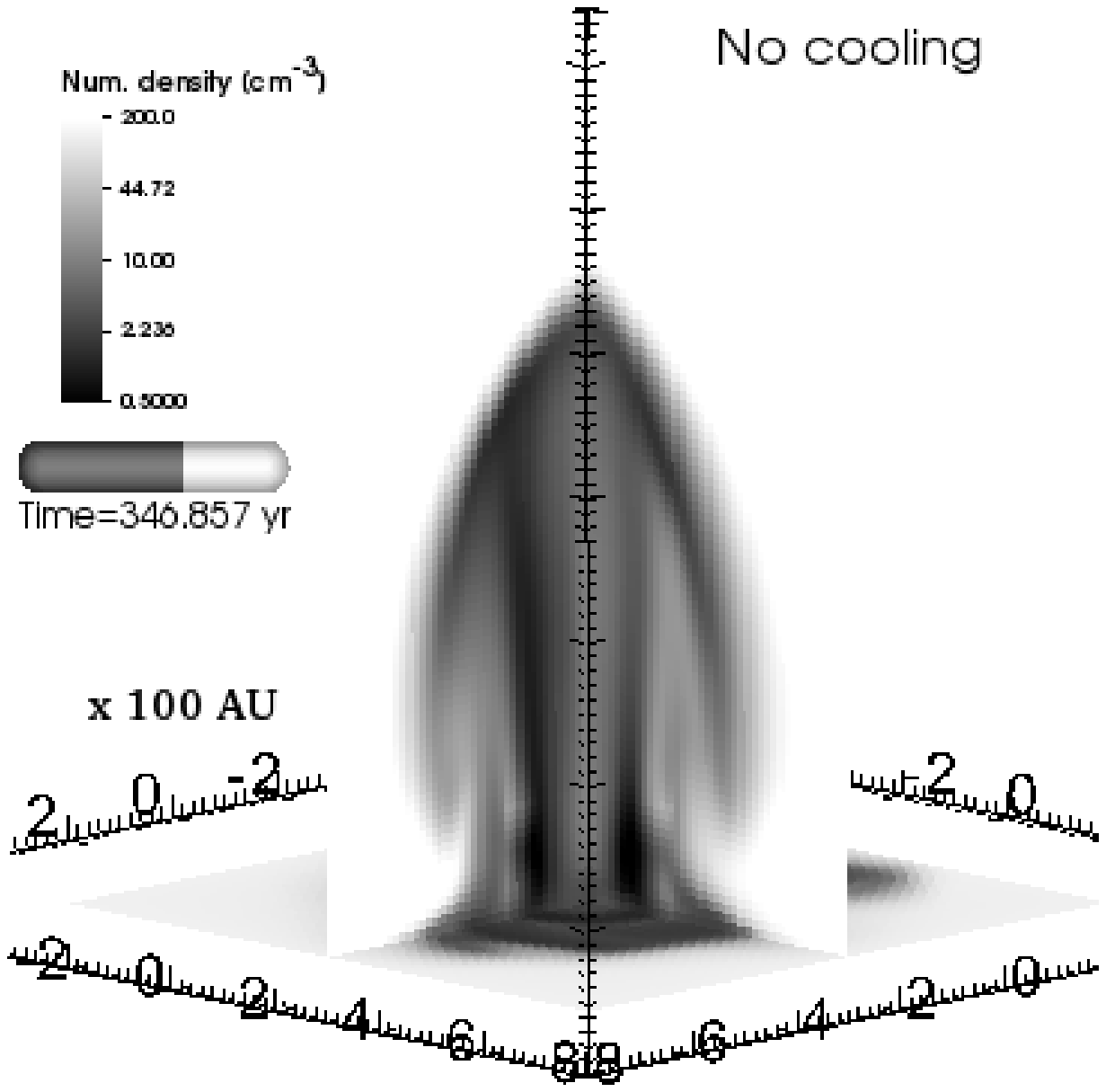} 
\includegraphics[width=.24\columnwidth]{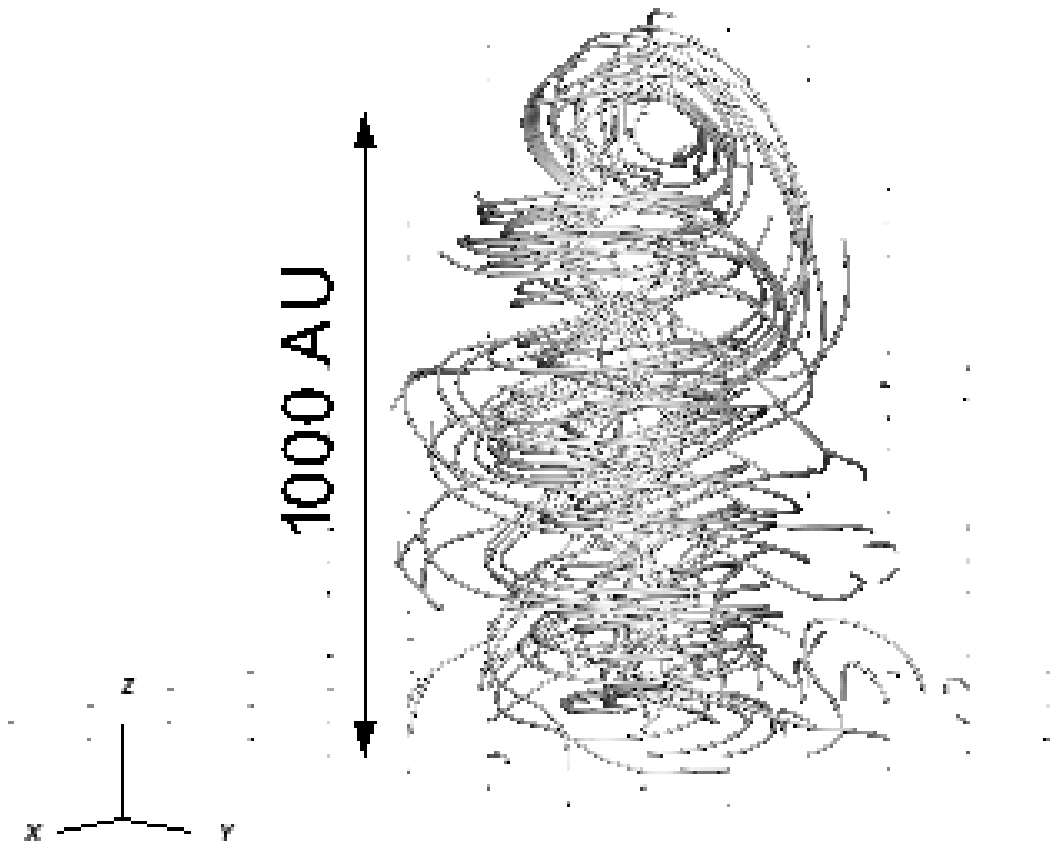} 
\includegraphics[width=.14\columnwidth]{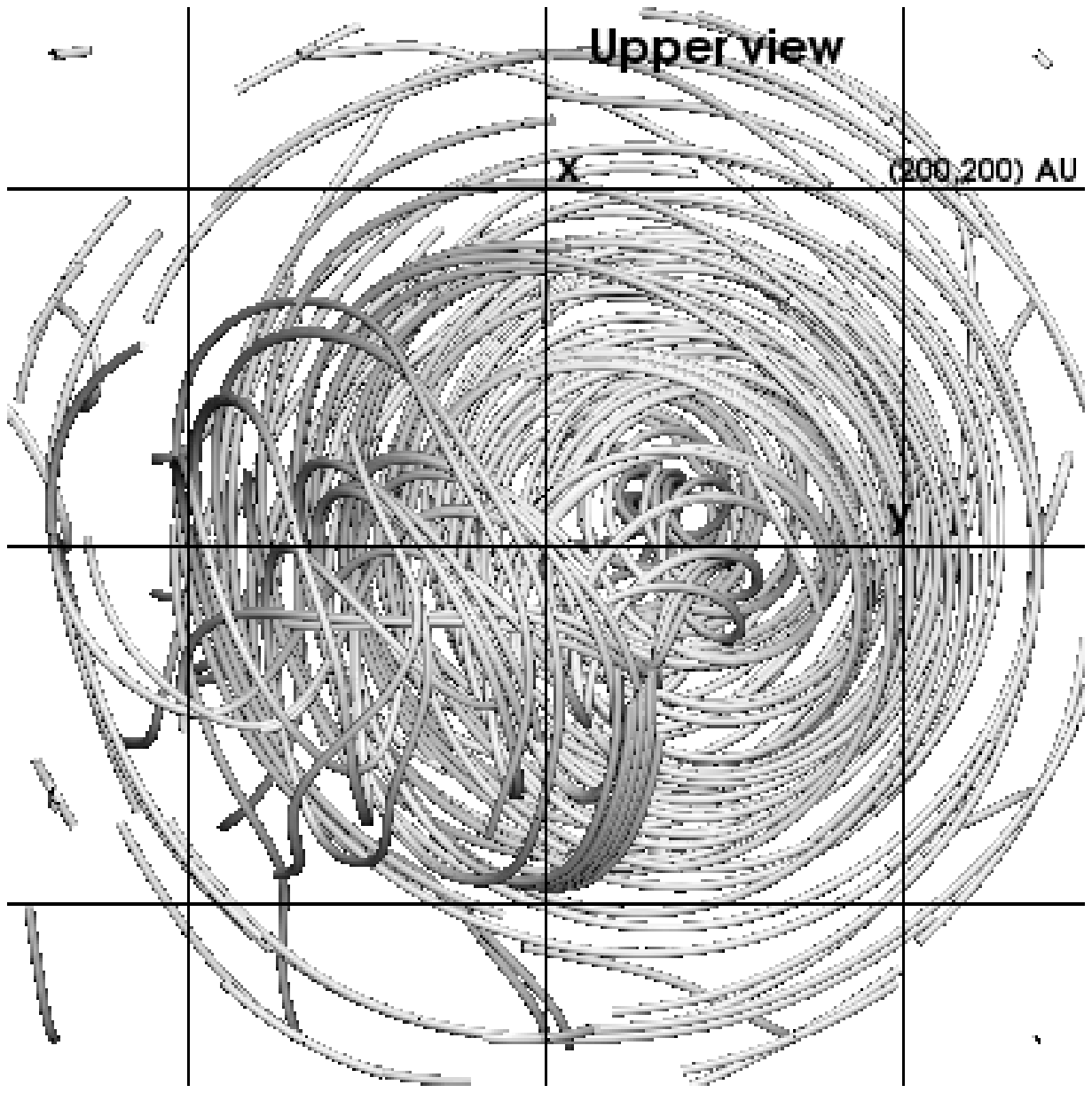} \\
\includegraphics[width=.24\columnwidth]{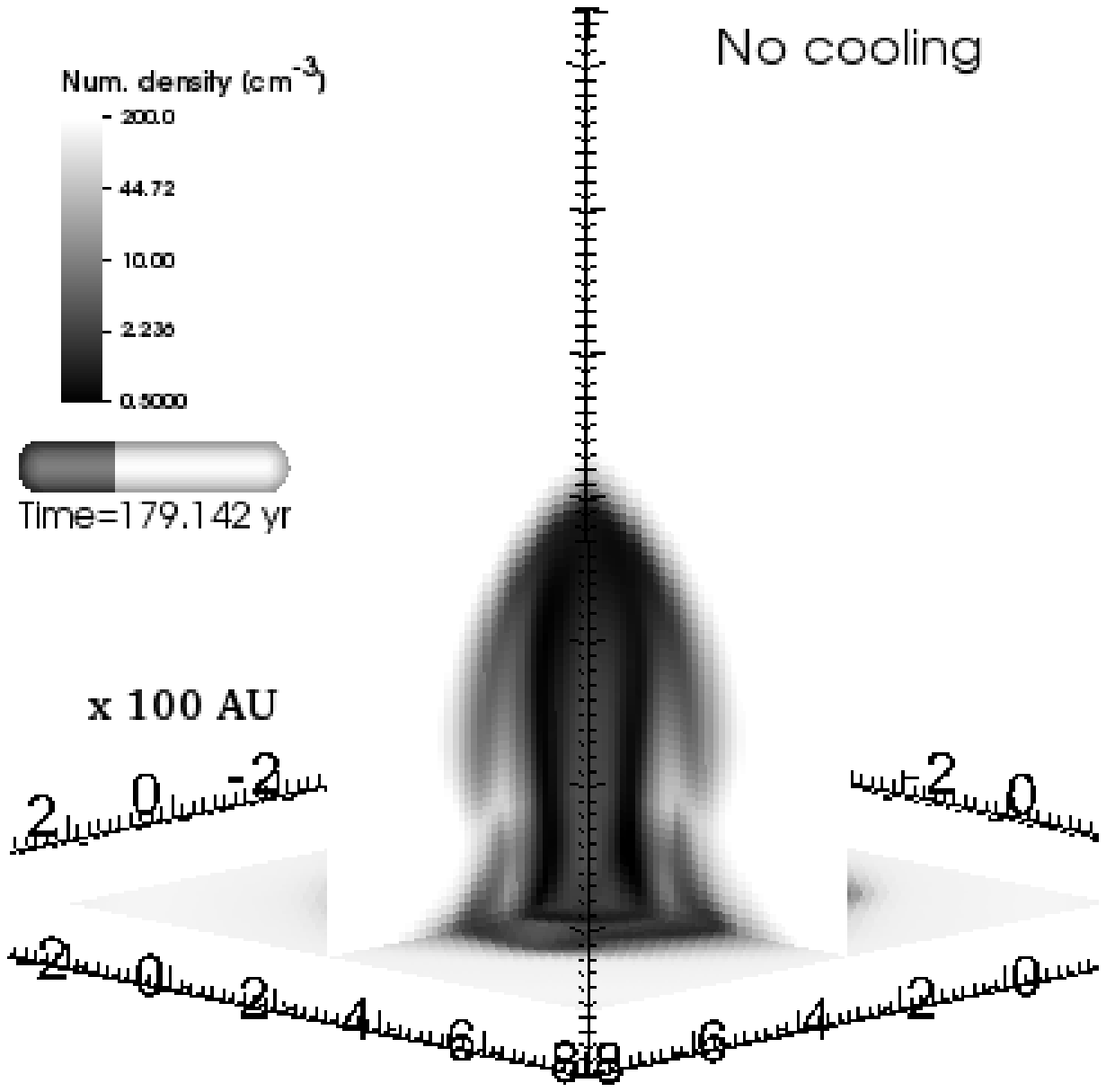} 
\includegraphics[width=.24\columnwidth]{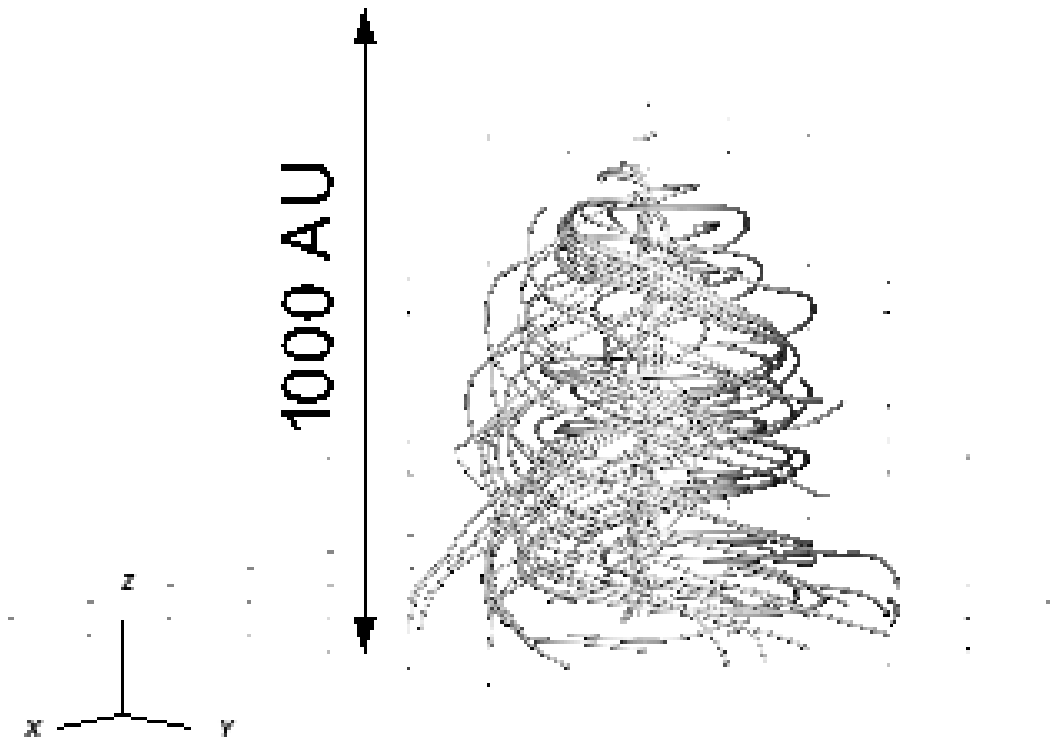} 
\includegraphics[width=.14\columnwidth]{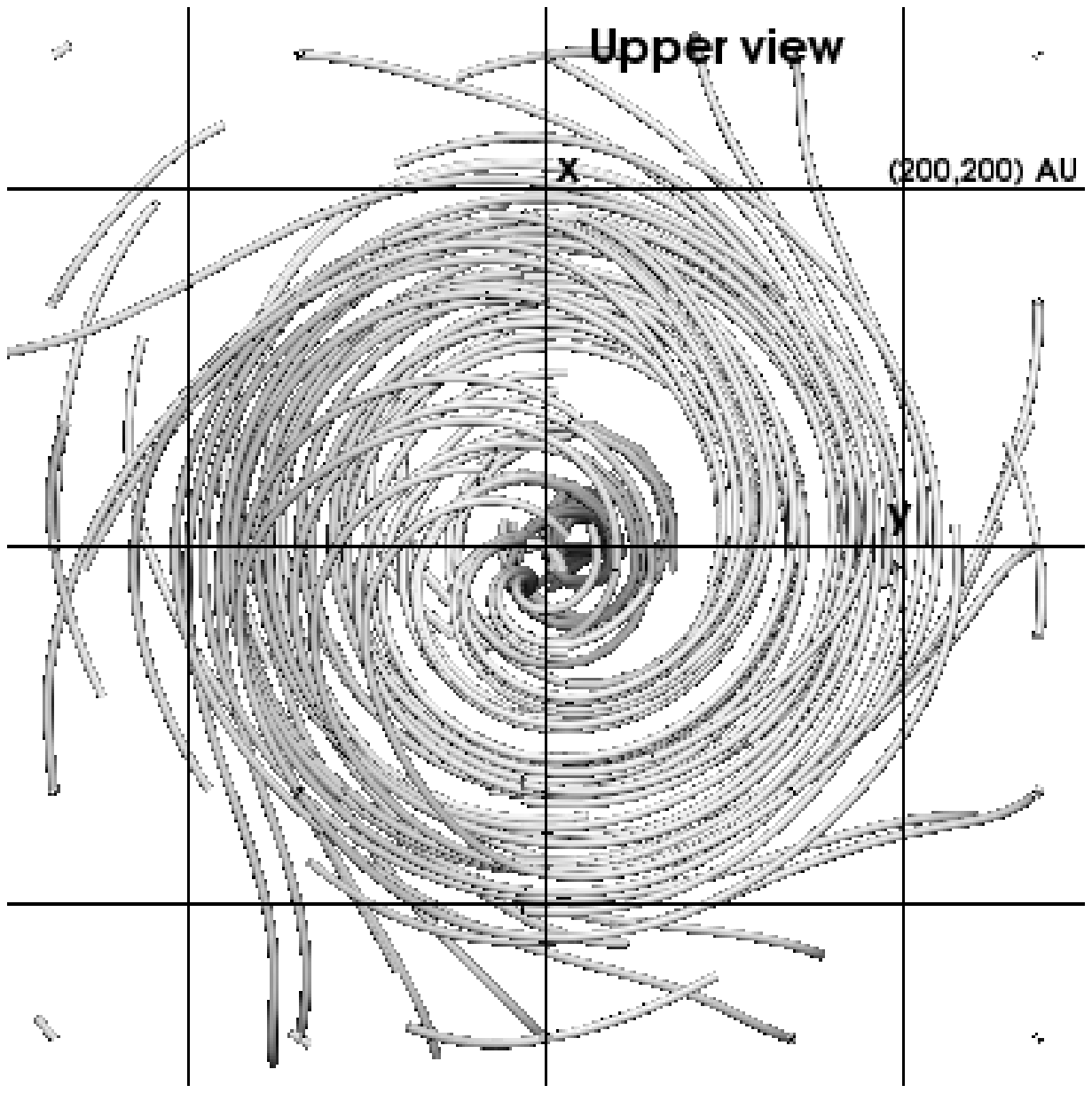} 
 \caption{Evolution of the adiabatic magnetic tower. Left: logarithmic density
grayscale maps. Middle: magnetic field lines. Right: field lines upper view.
Time is the same row-wise. Open field lines are a visualization effect. 
}
   \label{nocool}
\end{center}
\end{figure}

\begin{figure}[ht]
\begin{center}
\includegraphics[width=.24\columnwidth]{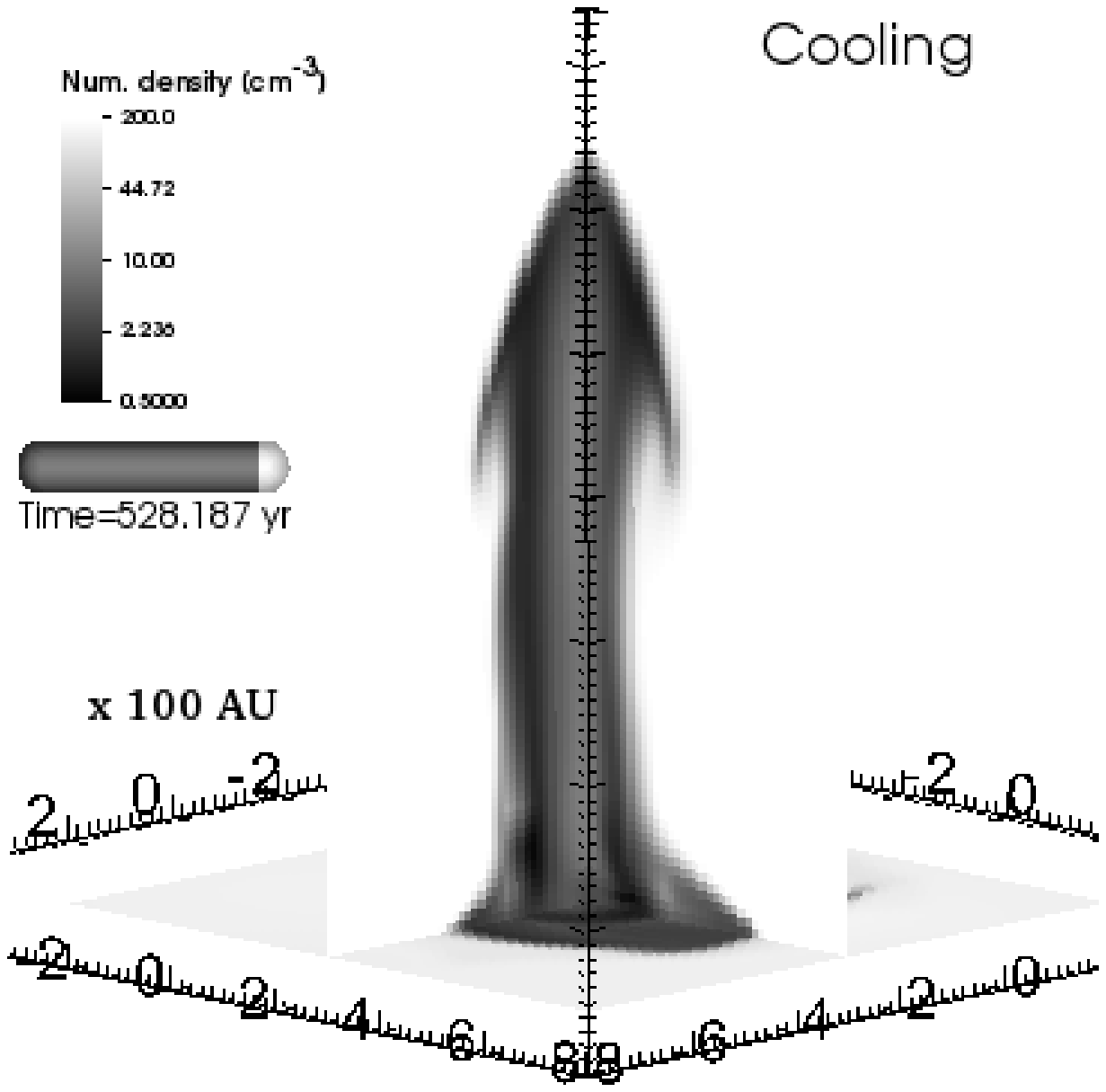} 
\includegraphics[width=.24\columnwidth]{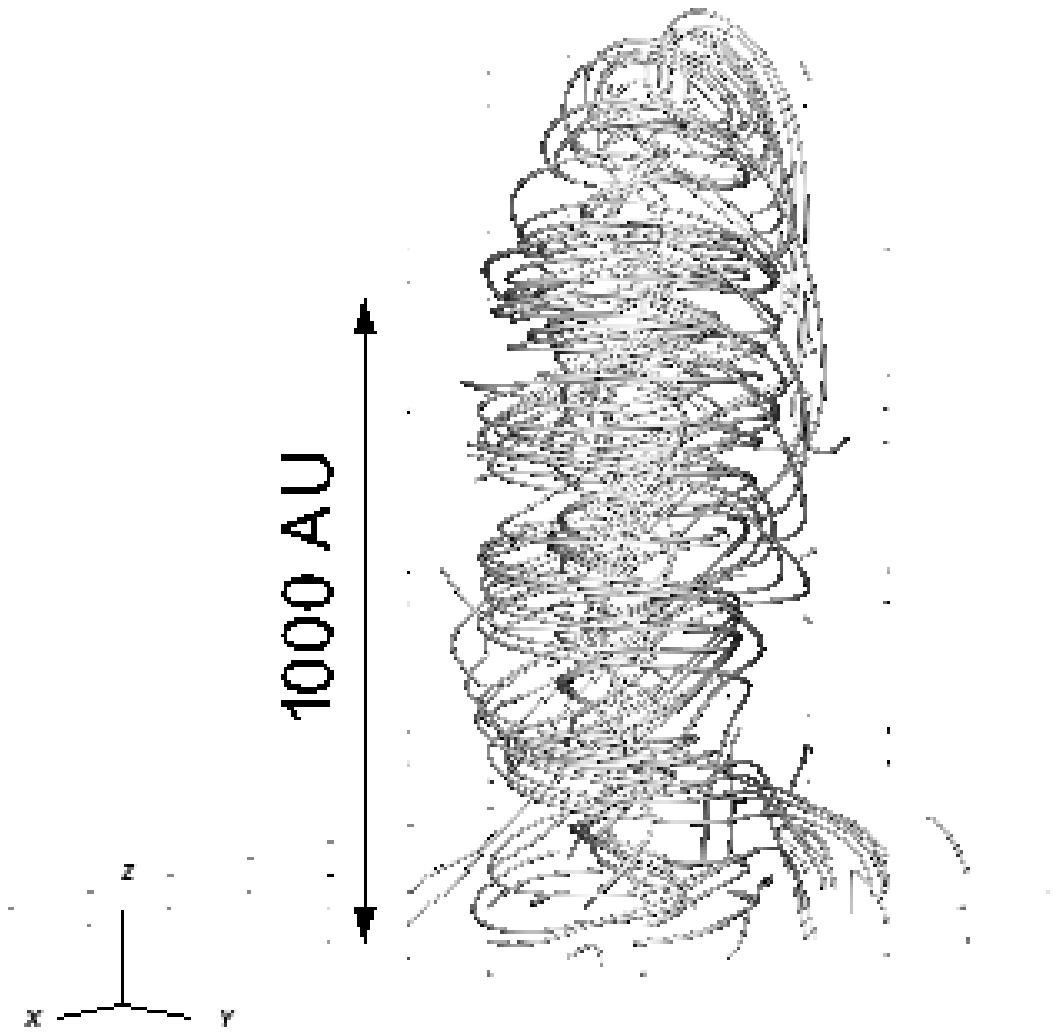} 
\includegraphics[width=.14\columnwidth]{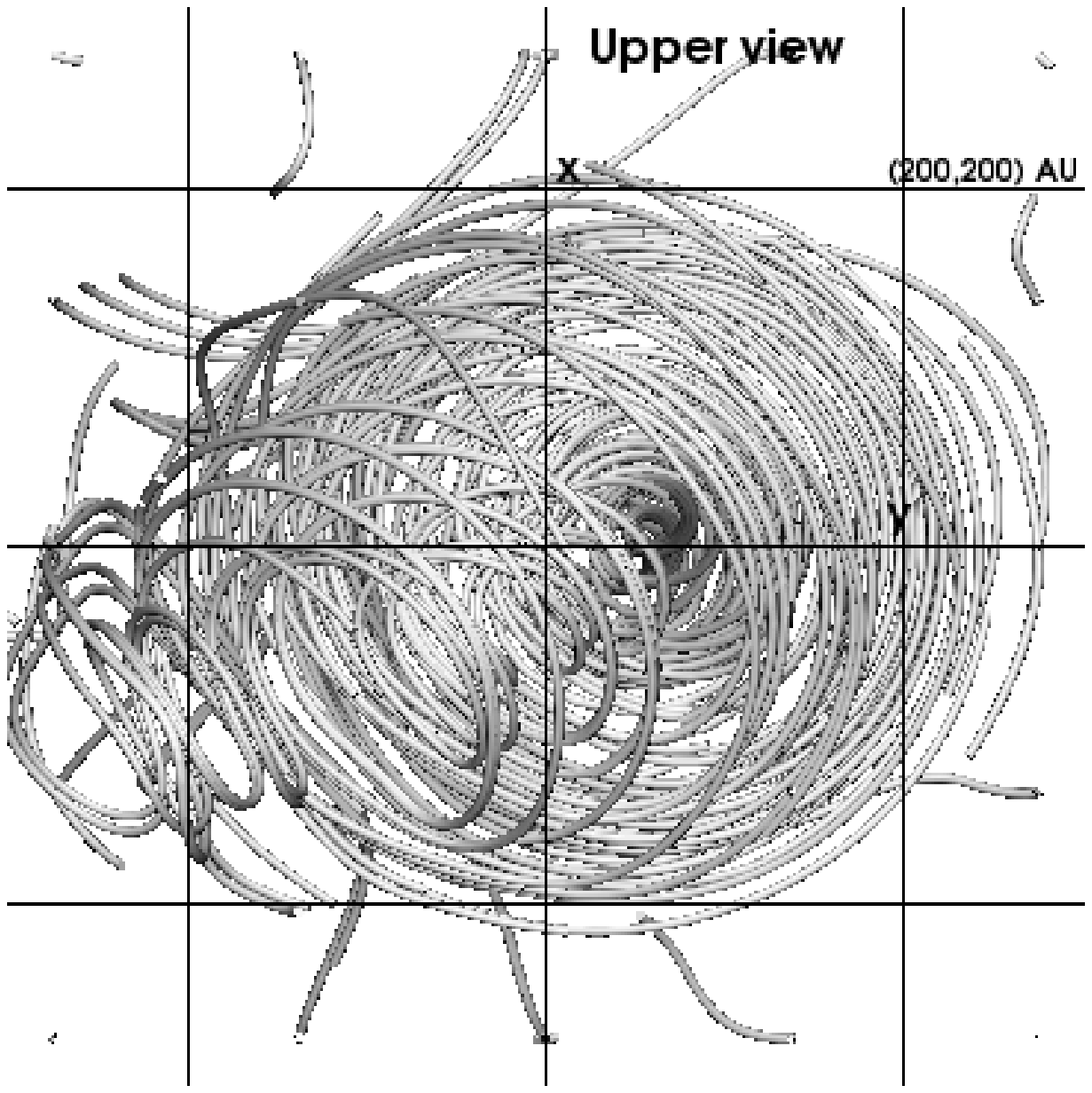} \\
\includegraphics[width=.24\columnwidth]{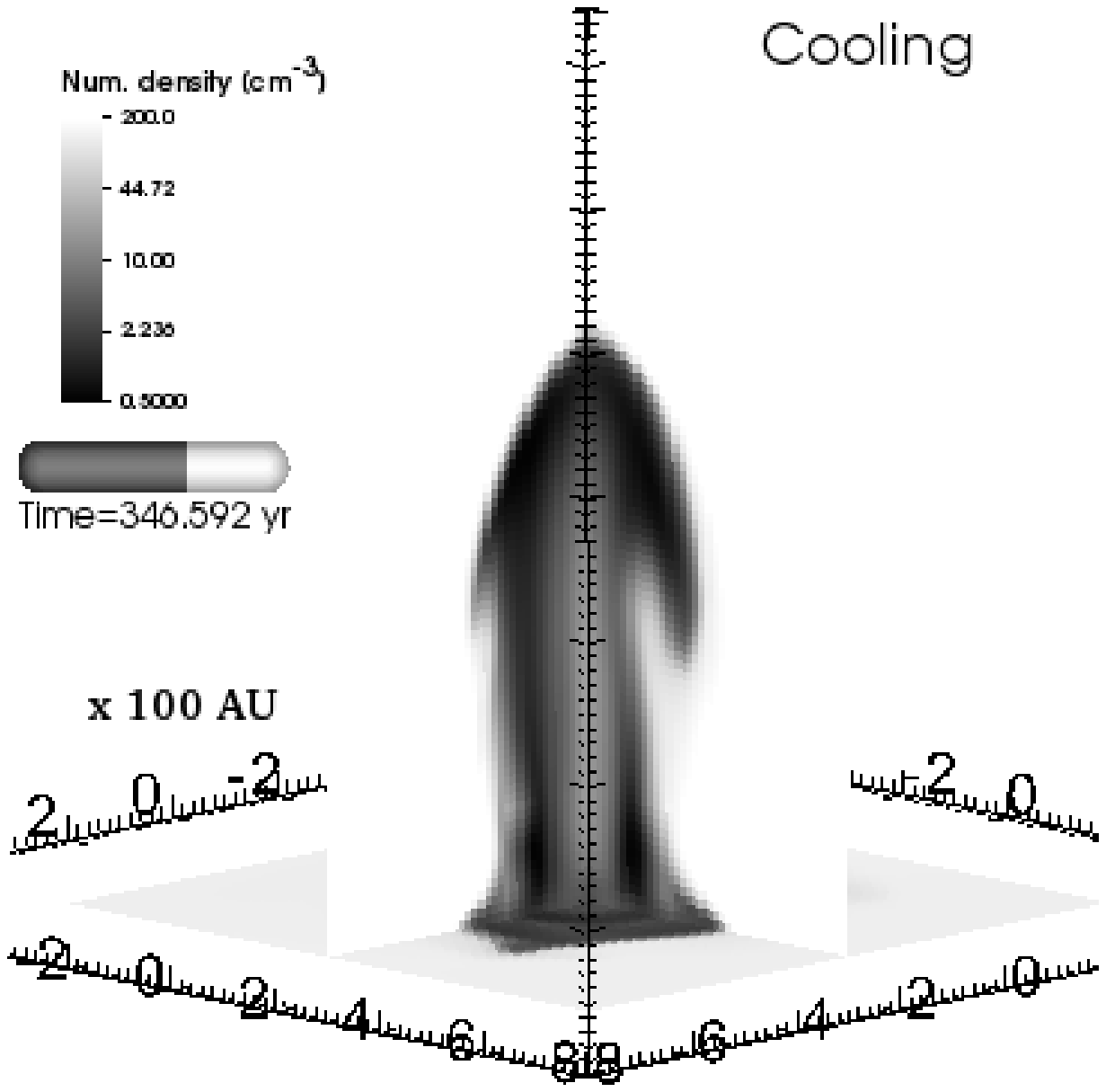} 
\includegraphics[width=.24\columnwidth]{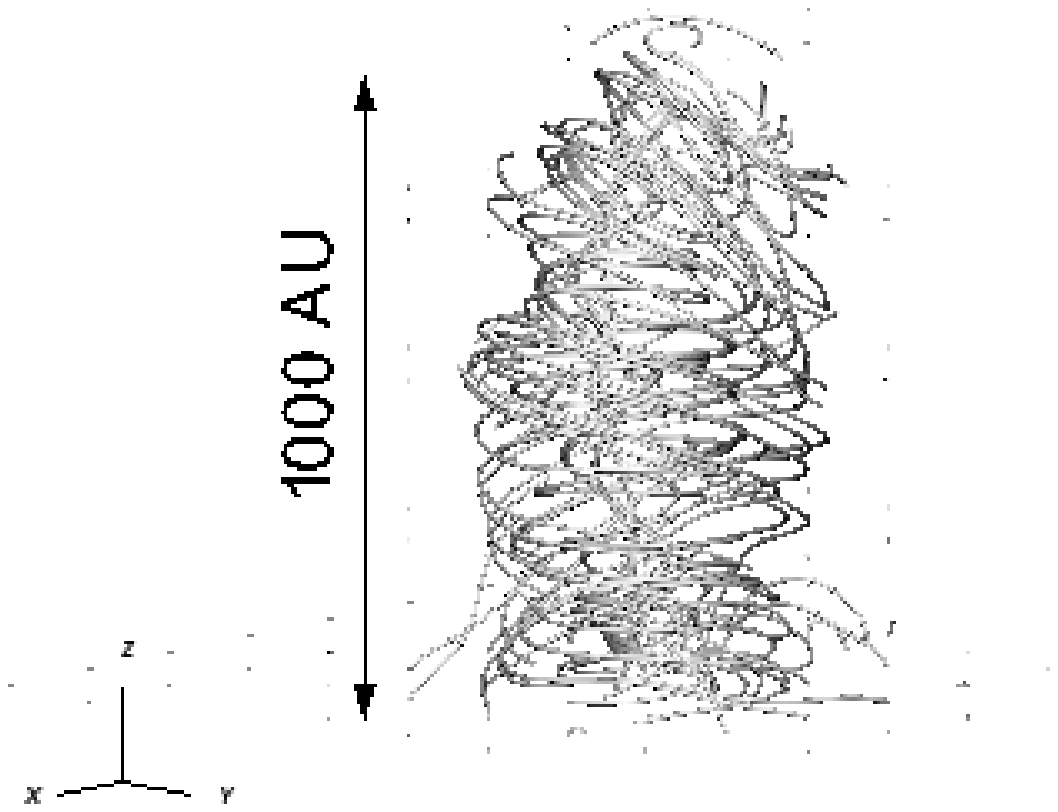} 
\includegraphics[width=.14\columnwidth]{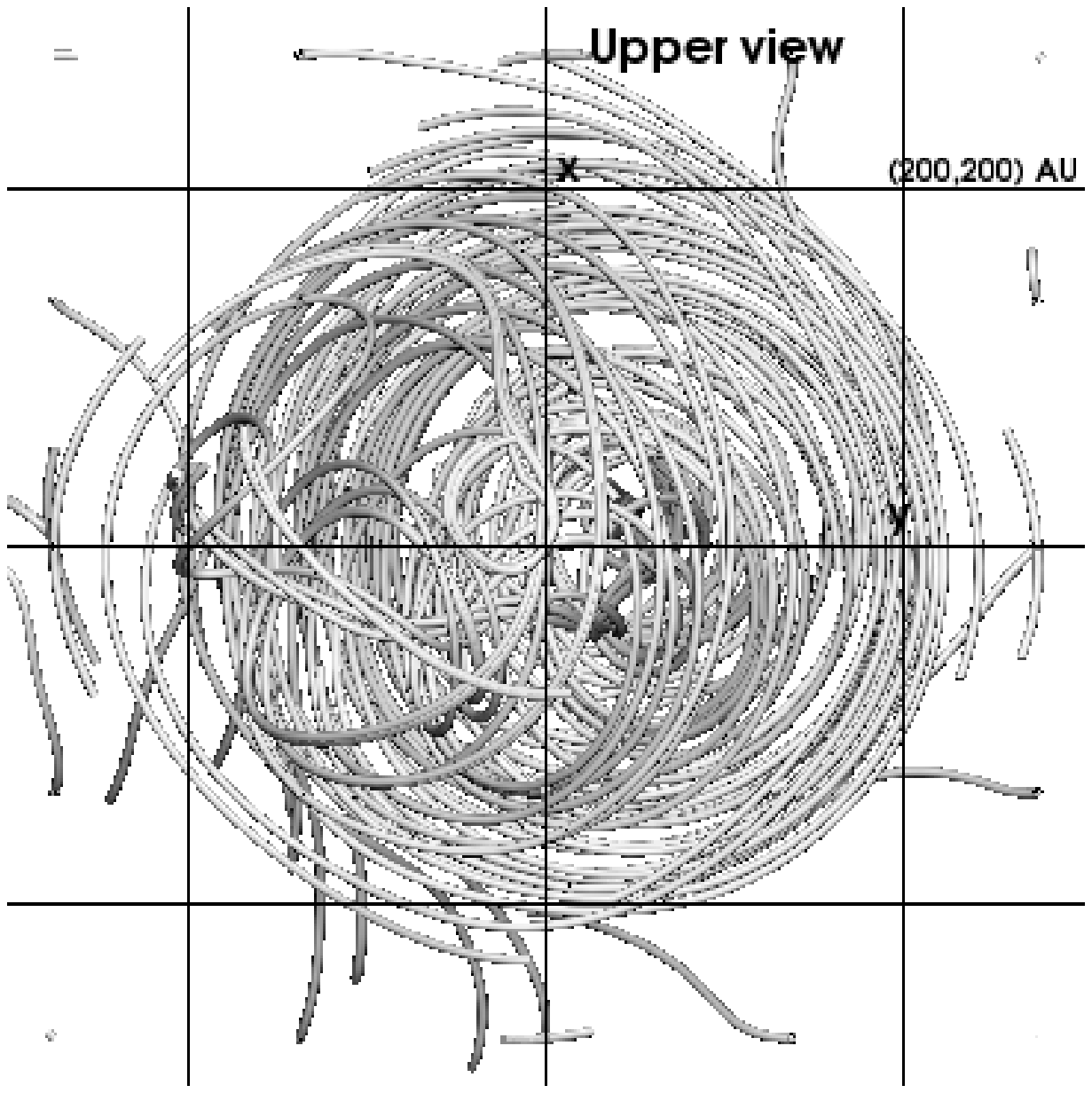} \\
\includegraphics[width=.24\columnwidth]{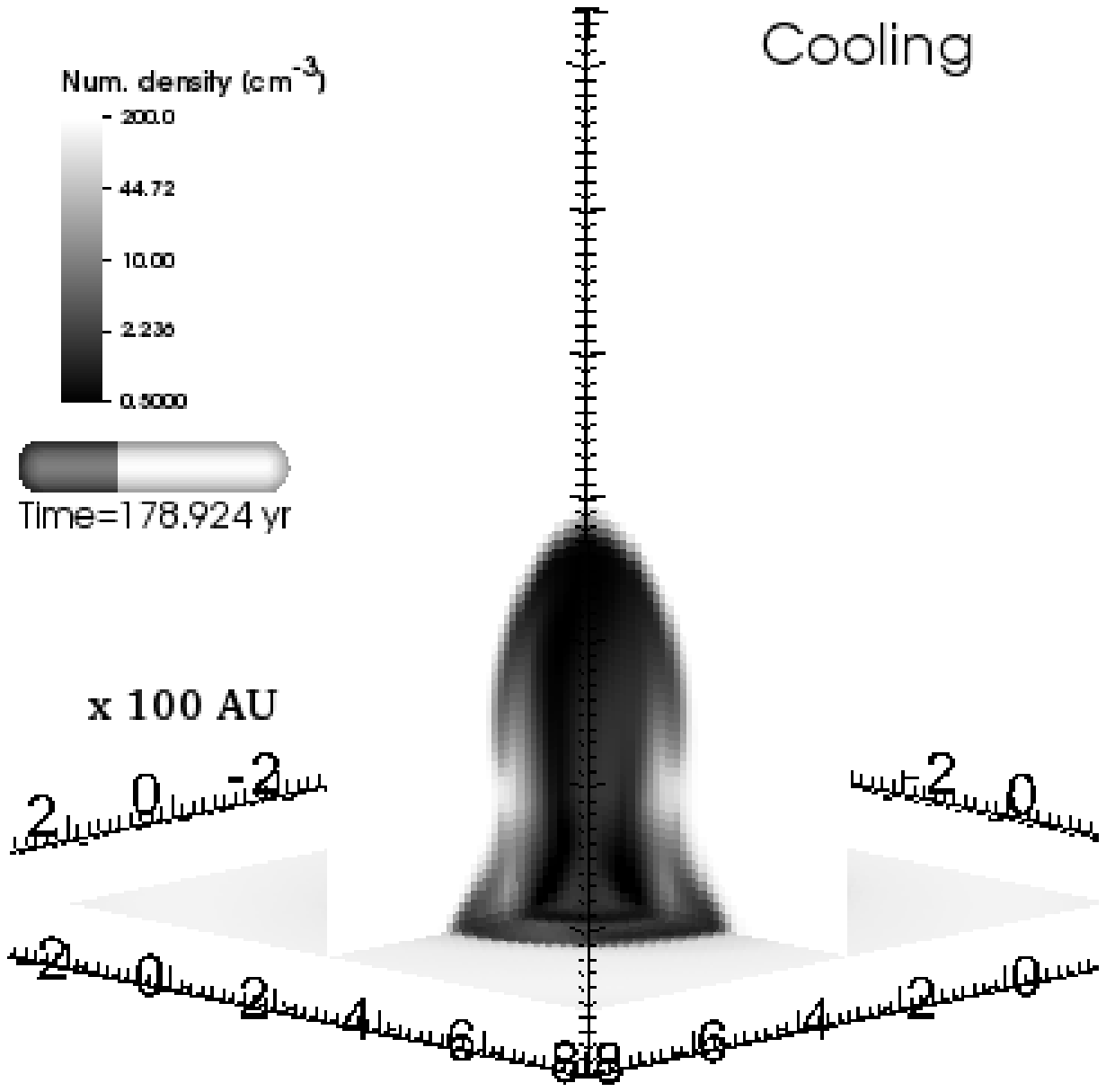} 
\includegraphics[width=.24\columnwidth]{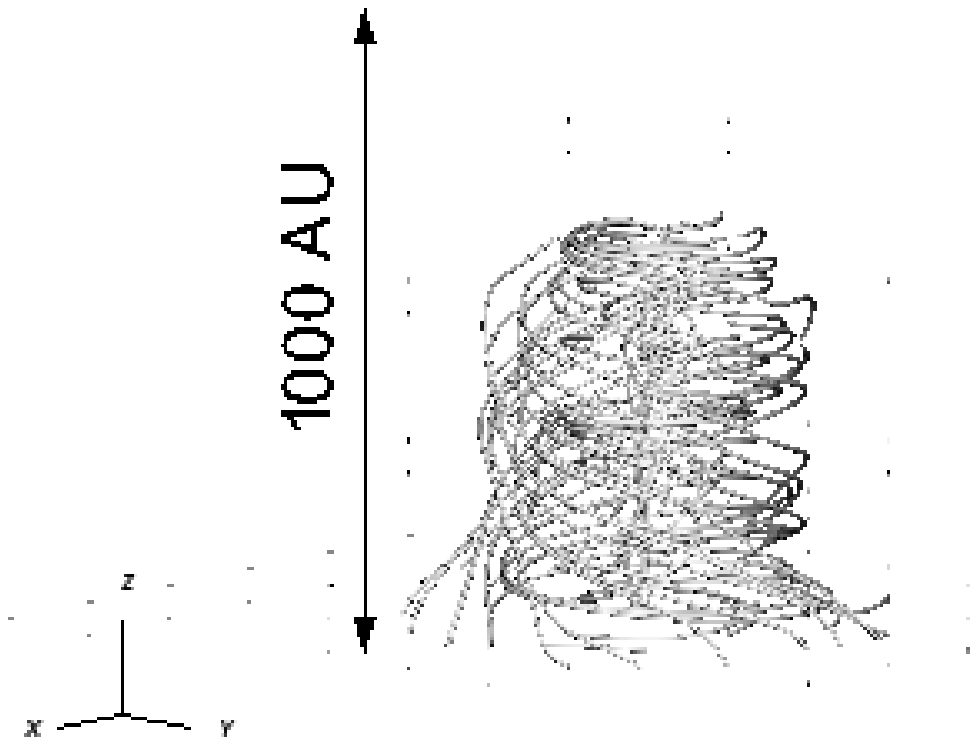} 
\includegraphics[width=.14\columnwidth]{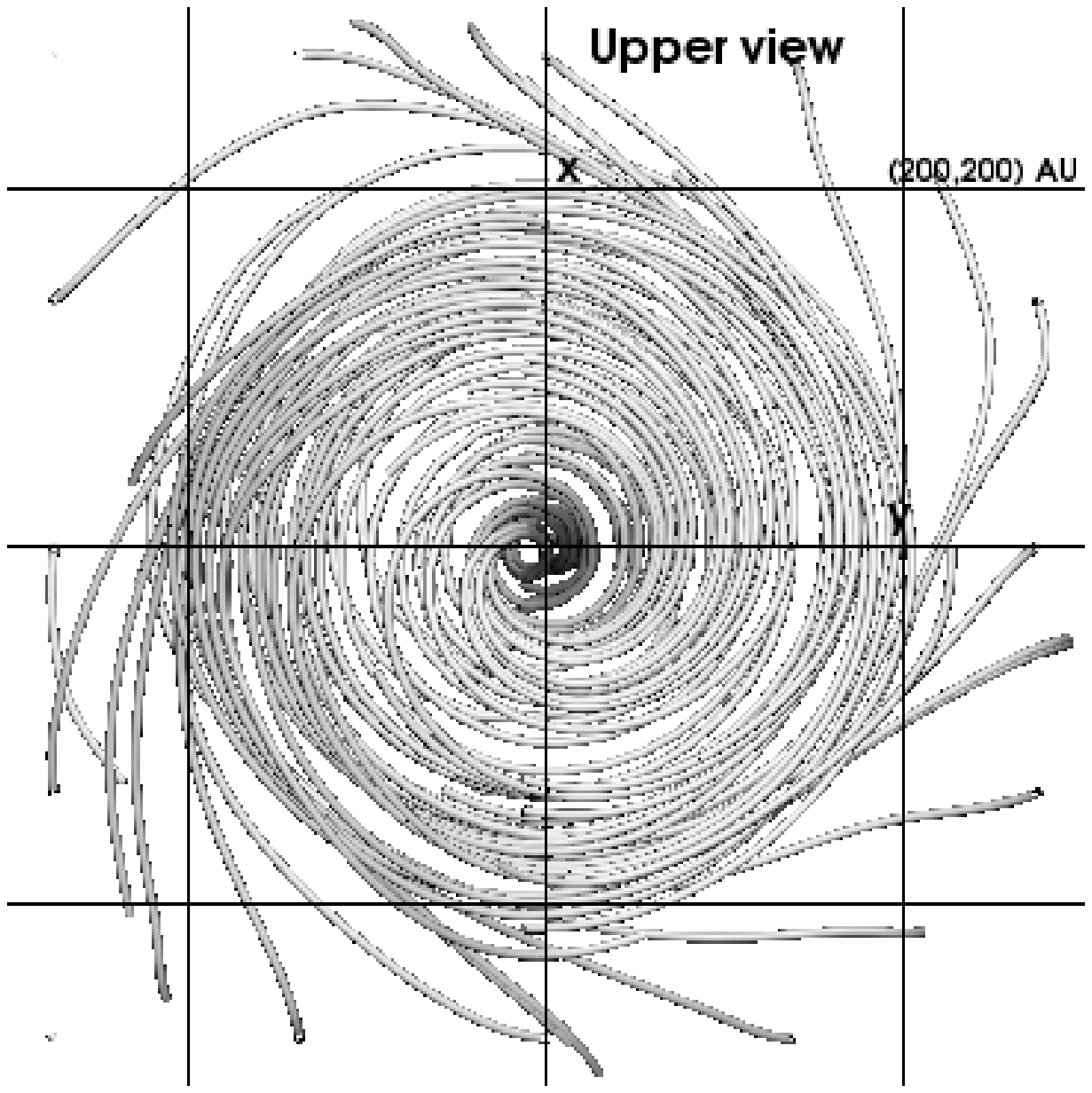} 
 \caption{Evolution of the cooling magnetic tower. Panel structure is as in
Figure~\ref{nocool}.
}
   \label{cool}
\end{center}
\end{figure}

The adiabatic and radiatively cooling magnetic towers are shown in
Figures~\ref{nocool} and~\ref{cool}, respectively.  The ratio of
thermal to magnetic pressures takes values within (0.1,1) at regions
where $r,z \leq\,$50$\pi$\,AU, throughout the simulations.
Vertical pressure gradients primarily due to the toroidal magnetic
field in the tower, cause the plasma to accelerate along the
$z$~direction, and a shock is driven in the plasma. A self-pinched
cavity \hbox{of $\sim\,$150\,AU} is formed in about~a~few~10\,yr by a
collimated outflow, the vertical speed of which
\hbox{is $\sim\,$100\,km\,s$^{-1}$}. This ``cocoon'' is filled with tightly
wound helical magnetic fields and gas that is about~0.01 times less
dense than that of the ambient. On average, the cocoon thermal
pressure is larger than the magnetic pressure by factors that increase
radially from~1 to~4, approximately, at $z \sim\,$300\,AU. This is
different than the magnetically dominated jets of \cite[Li~et~al.
(2006)]{li06}.  We note that, as opposed to other magnetized jet
launch simulations (e.g. \cite[Shibata \& Uchida 1986]{shibata86}),
no rotation has been imposed on the plasma.

In the cocoon, the toroidal magnetic flux  dominates the poloidal
flux. A poloidal electric current develops and the radial pressure
gradient from the toroidal magnetic field causes radial expansion
of the adiabatic tower (see left column of Figure~\ref{nocool})
that exceeds its collimating hoop stress.
 However, Figure~\ref{cool} shows that cooling
suppresses this effect significantly.  The upper view of the magnetic
field lines (rightmost column in Figures~\ref{nocool} and~\ref{cool})
shows two nested toroidal magnetic surfaces. The ambient pressure
collimates the outer magnetic surface, whereas its magnetic tension
collimates the field lines of the inner magnetic surface. Such
structures resemble the ones seen in the millimetric-scale magnetic
towers produced in laboratory experiments by \cite[Ciardi et al.
(2005, 7)]{lab1,2}.  Pinch instabilities cause the towers to bend,
and poloidal field lines seem to pile~up, particularly in the cooling
tower.

\begin{discussion}

ROMERO: In the case you have as a compact object a neutron star,
can you estimate the maximum field on the surace of the star necessary
to allow the formation of a tower? How high is this field? \\

HUARTE-ESPINOSA: I don’t know. I suppose I could because I can
control the injection of magnetic energy. \\

FERREIRA: YSO observations show that there is not enough pressure
to conine jets. So, could magnetic towers be formed or even maintained?
\\

HUARTE-ESPINOSA: The fields should expand. It could be interesting
to do simulations with other, weaker density profiles to see what
happens to the internal magnetic surface or jet. See the comment
by David Meier. \\

MEIER: Concerning the confinement question, manetic towers do not
need large external pressure to confine them- certainly nothing
like the pressure in the visible jet plasma. The main jet is confined
in the foward current part of the jet. Then moving radially outward,
the jet is self-confined in the current-free region. Finally, along
the return current surface (from the core of the jet) the magnetic
pressure is weak, and it only requires a weak ambient pressure to
confine the outer magnetic tower. \\

YUAN: I the jet is magnetically dominated, why cooling can play an
important dynamical role? \\

HUARTE-ESPINOSA: Cooling can play an important role in the shock
region at large distances from the central magnetized region. \\

FENDT: I expect reconnection being much more important than cooling
as the whole field structure will be reconfigured. What do you
think? \\

HUARTE-ESPINOSA: Reconnection plays a role in reality indeed. It
would be interesting to redo the simulations with magnetic resistivity
to see this. I honestly don’t know how important reconnection would
be relative to cooling. Yet, the towers in my simulations do not
expand due to rotation. So, line reconfiguration would not be as
important as it would for a rotation-formed tower.

\end{discussion}

\end{document}